\documentclass[twocolumn,english,aps,prl,superscriptaddress,nofootinbib]{revtex4-1}
\pdfoutput=1
\usepackage{graphicx}
\usepackage{amssymb,amsmath}
\usepackage{textcomp} 
\usepackage[bold]{hhtensor}
\usepackage{natbib}
\usepackage{lipsum}

\newcommand\blfootnote[1]{%
  \begingroup
  \renewcommand\thefootnote{}\footnote{#1}%
  \addtocounter{footnote}{-1}%
  \endgroup
}

\newcommand{\tocless}[2]{\bgroup\let\addcontentsline=\nocontentsline#1{#2}\egroup}

\begin{document}

\title{Nanophotonic lithium niobate electro-optic modulators}
\author{Cheng Wang*}
\affiliation{John A. Paulson School of Engineering and Applied Sciences, Harvard University, Cambridge, Massachusetts 02138, USA}
\author{Mian Zhang*}
\blfootnote{\vskip -0.2in *These authors contributed equally to this work}
\affiliation{John A. Paulson School of Engineering and Applied Sciences, Harvard University, Cambridge, Massachusetts 02138, USA}
\author{Brian Stern}
\affiliation{School of Electrical and Computer Engineering, Cornell University, Ithaca, New York 14853, USA.}
\affiliation{Department of Electrical Engineering, Columbia University, New York, New York 10027,USA}
\author{Michal Lipson}
\affiliation{Department of Electrical Engineering, Columbia University, New York, New York 10027,USA}
\author{Marko Loncar}
\affiliation{John A. Paulson School of Engineering and Applied Sciences, Harvard University, Cambridge, Massachusetts 02138, USA}
\date{\today}

\begin{abstract}
Modern communication networks require high performance and scalable electro-optic modulators that convert electrical signals to optical signals at high speed. Existing lithium niobate modulators have excellent performance but are bulky and prohibitively expensive to scale up. Here we demonstrate scalable and high-performance nanophotonic electro-optic modulators made of single-crystalline lithium niobate microring resonators and micro-Mach-Zehnder interferometers. We show a half-wave electro-optic modulation efficiency of 1.8V$\cdot$cm and data rates up to 40 Gbps.
\end{abstract}
\maketitle

\date{\today}
\newcommand{\nocontentsline}[3]{}

Data centers, metropolitan and long-haul data communication networks demand scalable and high-performance electro-optic modulators to convert electrical signals to modulated light waves at high speed \cite{Wooten00,Janner09}. For decades, lithium niobate (LiNbO$_3$, LN) has been the material of choice owing to its excellent properties - namely large electro-optic response, high intrinsic bandwidth, wide transparency window, exceptional signal quality and good temperature stability \cite{Wooten00,Janner09,Nikogosyan05}. Existing LN modulators however are not scalable due to the difficulty in nanostructuring LN \cite{Janner09}. As a result, they remain bulky ($\sim$ 10 cm long), discrete, expensive, and require high-power electrical drivers \cite{Wooten00,Janner09}. Integrated silicon (Si) \cite{Xu05,Timurdogan14,Zhang16,Sun15} and indium phosphide (InP) \cite{Rolland93,Aoki93,Kikuchi12} photonics are promising solutions for scalability but come at the cost of compromised performance \cite{Xu05,Timurdogan14,Zhang16,Sun15,Rolland93,Aoki93,Kikuchi12}. Here we demonstrate monolithically integrated LN electro-optic modulators that are orders of magnitudes smaller and more efficient than traditional bulk LN devices while preserving LN's excellent material properties. Our compact LN platform consists of low-loss nanoscale LN waveguides, micro-ring resonators and miniaturized Mach-Zehnder interferometers (MZI), fabricated by directly shaping LN thin films into sub-wavelength structures. 

To address the need of scalable and high-performance electro-optic modulators for modern data communications, tremendous efforts have been made in recent years towards a variety of platforms that feature small footprints and high data bandwidths, including Si \cite{Xu05,Timurdogan14,Zhang16,Sun15}, InP \cite{Rolland93,Aoki93,Kikuchi12}, AlN \cite{Xiong12}, plasmonics \cite{Haffner15}, graphene \cite{Liu11,Phare15,Novoselov12} and polymers \cite{Lee02,Clark10}. However, these approaches have fundamental material limitations that are hard to overcome. For example, both Si and InP modulators rely on switching mechanisms (carrier injection and quantum-confinement Stark effect respectively) that are intrinsically nonlinear, absorptive, and sensitive to temperature fluctuation \cite{Xu05,Timurdogan14,Zhang16,Sun15,Rolland93,Aoki93,Kikuchi12}. Limitations of the other platforms include low switching efficiency (AlN) \cite{Xiong12}, high optical loss (plasmonics) \cite{Haffner15}, challenging scalability (graphene) \cite{Novoselov12} and poor long term stability (polymer) \cite{Clark10}. 

\begin{figure*}
   \centering
   \includegraphics[angle=0]{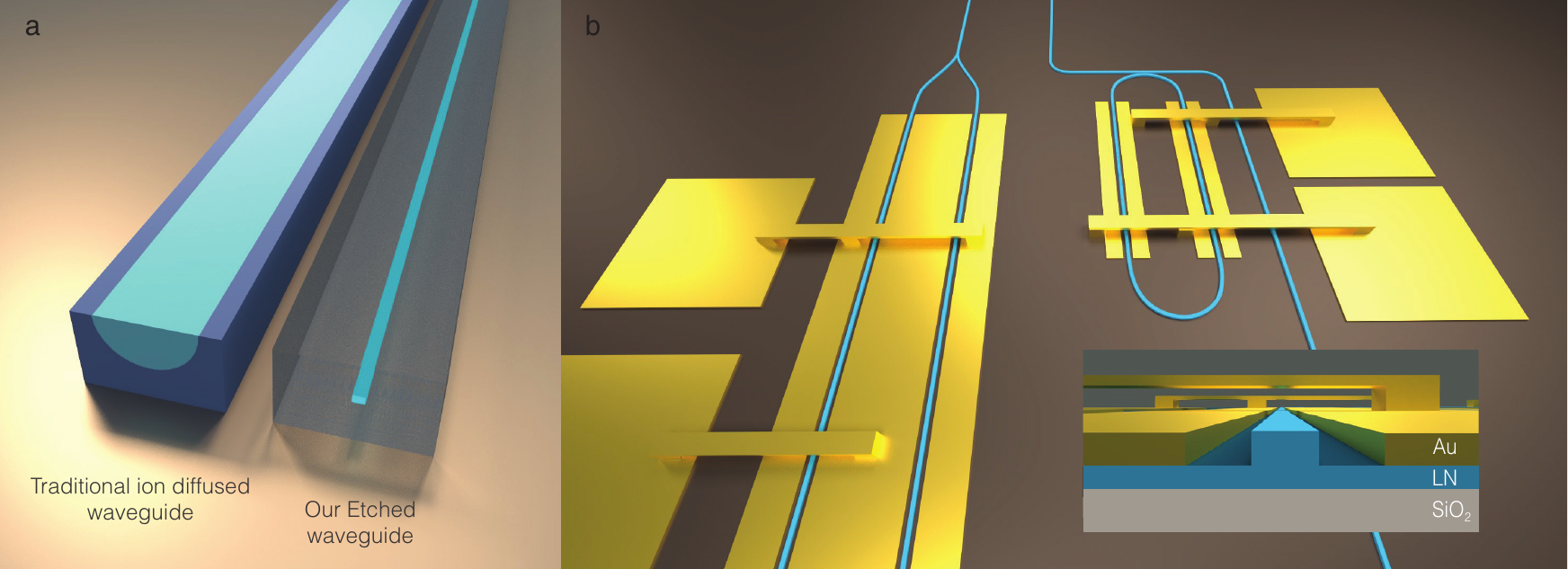}

 \caption{\label{fig1}\textbf{Device design.} \textbf{a}. Schematics of a traditional ion-diffused LN waveguide (left) and our etched LN waveguide embedded in SiO$_2$ (right), roughly to scale. The light blue region indicates the approximate waveguiding core in each circumstance. The larger index contrast in etched waveguides allows for much stronger light confinement. \textbf{b}. Schematic view of the device layout with thin film LN waveguides and RF electrodes. Metal vias and bridges are fabricated to achieve modulation on both sides of the devices. Inset shows a schematic of the device cross-section with an overview of the metal bridge.}

 \end{figure*}

\begin{figure*}
   \centering
   \includegraphics[angle=0]{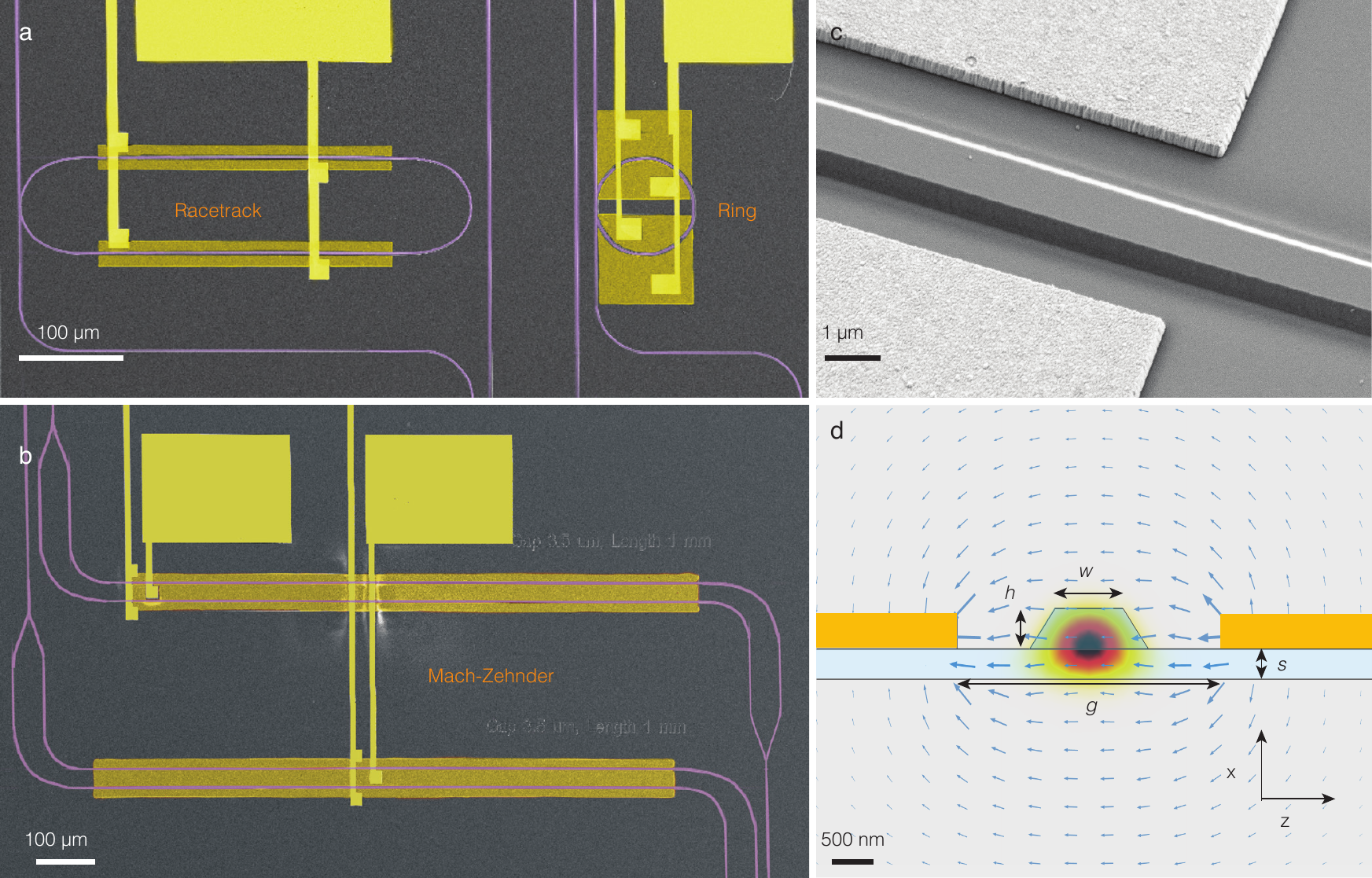}

 \caption{\label{fig2}\textbf{Fabricated optical devices and electrical contacts} \textbf{a-b}. False-color scanning electron microscope (SEM) images of the fabricated racetrack and ring resonator based modulators (a) and Mach-Zehnder interferometer based modulators (b). \textbf{c}. Cross-section view of the simulated optical TE mode profile (E$_z$ component) and RF electrical field (shown by arrows). The $x$-cut LN used here is most sensitive to the horizontal component of the electric field (E$_z$). $h$: LN waveguide height; $w$: waveguide width; $s$: LN slab thickness. $g$: metal electrode gap. \textbf{d}. Close-up SEM image of the metal electrodes and the optical waveguide.}

\end{figure*}

LN remains the material of choice for high-performance electro-optic modulators due to its wide bandgap (high transparency) and large second order ($\chi_2$) electro-optic coefficient (30 pm/V). In contrast to Si and InP, the $\chi_2$ process in LN changes its index of refraction linearly with an applied electrical field, at femtosecond timescale. The efficiency of this process is determined by the overlap of the optical and the electrical fields. Conventional ion-diffused LN waveguides suffer from the low refractive index contrast ($\Delta n < 0.02$) between core and cladding (Fig. \ref{fig1}a), resulting in large optical modal volumes and bending radii \cite{Schmidt74}. As a result, the photonic structures are large and the radio-frequency (RF) electrodes have to be placed far away from the optical mode to prevent detrimental waveguide propagation loss, significantly reducing the electro-optic switching efficiency \cite{Wooten00,Janner09}.

\begin{figure*}[t!]
   \centering
   \includegraphics[angle=0]{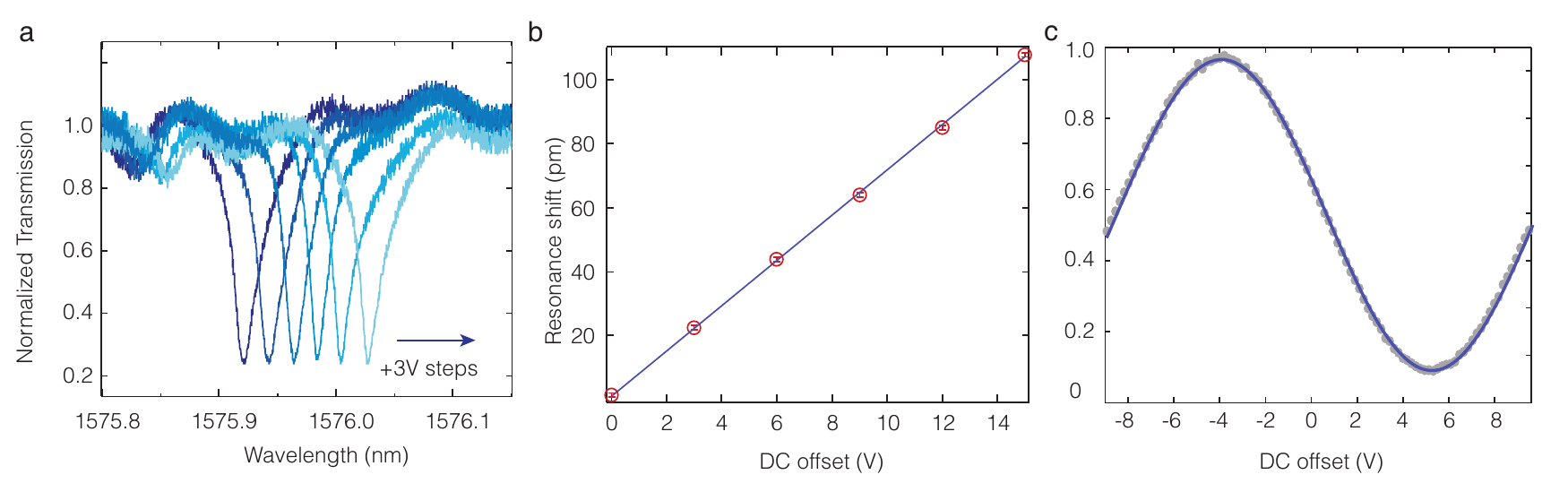}

 \caption{\label{fig3}\textbf{DC electrical and optical characterization} \textbf{a}. Measured transmission spectra of a high $Q$ ($\sim$ 50,000) racetrack resonator exhibits large frequency shift with applied DC voltages. \textbf{b}. Linear resonant wavelength shift as a function of DC voltage with error bars. The measured tuning efficiency is 7.0 pm/V. \textbf{c}. Optical transmission of a 2 mm long MZI modulator versus DC voltage applied, indicating a half-wave voltage ($V_\pi$) of 9 V and a voltage-length product of 1.8 V$\cdot$cm.}
\end{figure*}

Thin-film LN on insulator has recently emerged as a promising candidate to shrink the optical modal volume and boost the electro-optic efficiency \cite{Poberaj12}. While progress has been made towards chip-scale LN devices \cite{Poberaj12,Guarino07,Chen14,Rao15,Jin16,Rao16,Wang15,Cai16,Wang14,Wang16,Chang16}, realization of a monolithic LN nanophotonic circuit remains challenging due to the difficulty in LN dry etching. Instead, heterogeneous integration of intact LN thin films with another easy-to-etch material has been pursued \cite{Chen14,Rao15,Jin16,Rao16}. These hybrid platforms, however, suffer from low bandwidth performance and/or low optical confinement due to the addition of non-ideal deposited materials. For example, the hybrid Si/LN modulator has a bandwidth of $\sim$ 5 GHz due to the large resistance of the bonded silicon layer \cite{Chen14}. Ridge waveguides formed by depositing passive index-matching materials (e.g. silicon nitride and chalcogenide) results in weakly confined optical modes and large bending ($>$ 0.2 mm) \cite{Rao15,Jin16,Rao16}.Furthermore, the electro-optic efficiencies in these heterogeneous devices are also compromised due to the reduced overlap between the electric field and the active material region \cite{Chen14,Rao15,Jin16,Rao16}.

We demonstrate single-crystalline LN photonic structures with submicron optical confinement, small bending radii ($<$ 20 $\mu$m) and low propagation loss by directly shaping single-crystalline LN into nanoscale waveguides (Fig. \ref{fig1}a). The waveguides are defined on thin-film LN-on-insulator substrates using electron beam lithography and subsequently dry etched in Ar$_+$ plasma using a deposited Si hard mask. The index contrast between the LN core and the silicon dioxide (SiO$_2$) cladding is $\Delta n = 0.67$, over an order of magnitude higher than ion-diffused LN waveguides. Figures \ref{fig2}a and \ref{fig2}b show a range of fabricated nanophotonic LN devices including nano-waveguides, ring resonators, racetrack resonators and MZIs. The typical propagation loss of these structures is $\sim$ 3 dB/cm, which is limited by etching roughness \cite{Wang14} and can be further improved by at least an order of magnitude \cite{Wang15,Ilchenko04}. The resulting MZI and racetrack structures have low on-chip insertion loss of $\sim$ 2 dB and $\sim$ 1 dB respectively (with additional $\sim$ 5 dB/facet coupling loss).

The highly confined optical mode allows us to maximize electro-optic modulation efficiency by placing gold micro-RF electrodes close to the LN waveguide (Figs. \ref{fig2}c and \ref{fig2}d). Our devices make use of an $x$-cut LN configuration, where transverse-electric (TE) optical modes and in-plane electric fields ($E_z$) interact through the highest electro-optic tensor component ($r_{33}$) of LN. We design the waveguide geometry and the micro-RF electrode positions to achieve the optimal overlap between the optical and electric fields, while minimizing the bending loss and the metal-induced absorption loss. Figure \ref{fig2}c shows the numerically simulated overlap between the corresponding optical and electric fields. The optical waveguides have a top width $w$ = 900 nm, rib height $h$ = 400 nm, and a slab thickness $s$ = 300 nm (Fig. \ref{fig2}c). To maximize the in-plane electric field ($E_z$), we sandwich the optical waveguide between the signal and ground electrodes with a gap of $g$ = 3.5 $\mu$m. A SiO$_2$ cladding layer is used to further enhance this overlap by increasing the dielectric constant of the surrounding media to match the high dielectric constant of LN ($\sim$ 28)\cite{Nikogosyan05}.

We show efficient and linear electro-optic tuning in a racetrack modulator and a micro-MZI modulator. Figure \ref{fig3}a shows a typical transmission spectrum of a racetrack resonator with a loaded quality ($Q$) factor $\sim$ 50,000. When a voltage is applied, the change of refractive index modifies the effective optical path length of the resonator, resulting in a resonance frequency shift. The electrical fields on the two racetrack arms are aligned to the same direction so that the modulation on the two arms adds up (Fig. \ref{fig1}). The measured electro-optic efficiency is 7.0 pm/V with good linearity and no observable changes in resonance extinction ratio and linewidth (Figs. \ref{fig3}a and \ref{fig3}b). The MZI modulator is a balanced interferometer with two 50:50 Y-splitters and two optical paths. The applied voltage induces a phase delay on one arm and a phase advance on the other, which in turn change the output intensity at the Y-combiner by interference. The minimum voltage that is needed to completely switch the output between on and off is defined as the half-wave voltage ($V_{\pi}$). We measure a $V_{\pi}$ of 9 V from a 2 mm long MZI modulator, with 10 dB extinction ratio (Fig. \ref{fig3}c). This translates to a voltage-length product of 1.8 V$\cdot$cm, an order of magnitude better than bulk LN devices \cite{Wang15,Ilchenko04} and significantly higher than previously reported LN thin-film devices because of the highly-confined electro-optic overlap \cite{Guarino07,Chen14,Rao15,Jin16,Rao16,Wang15,Cai16}.

\begin{figure*}
     \centering
     \includegraphics[angle=0]{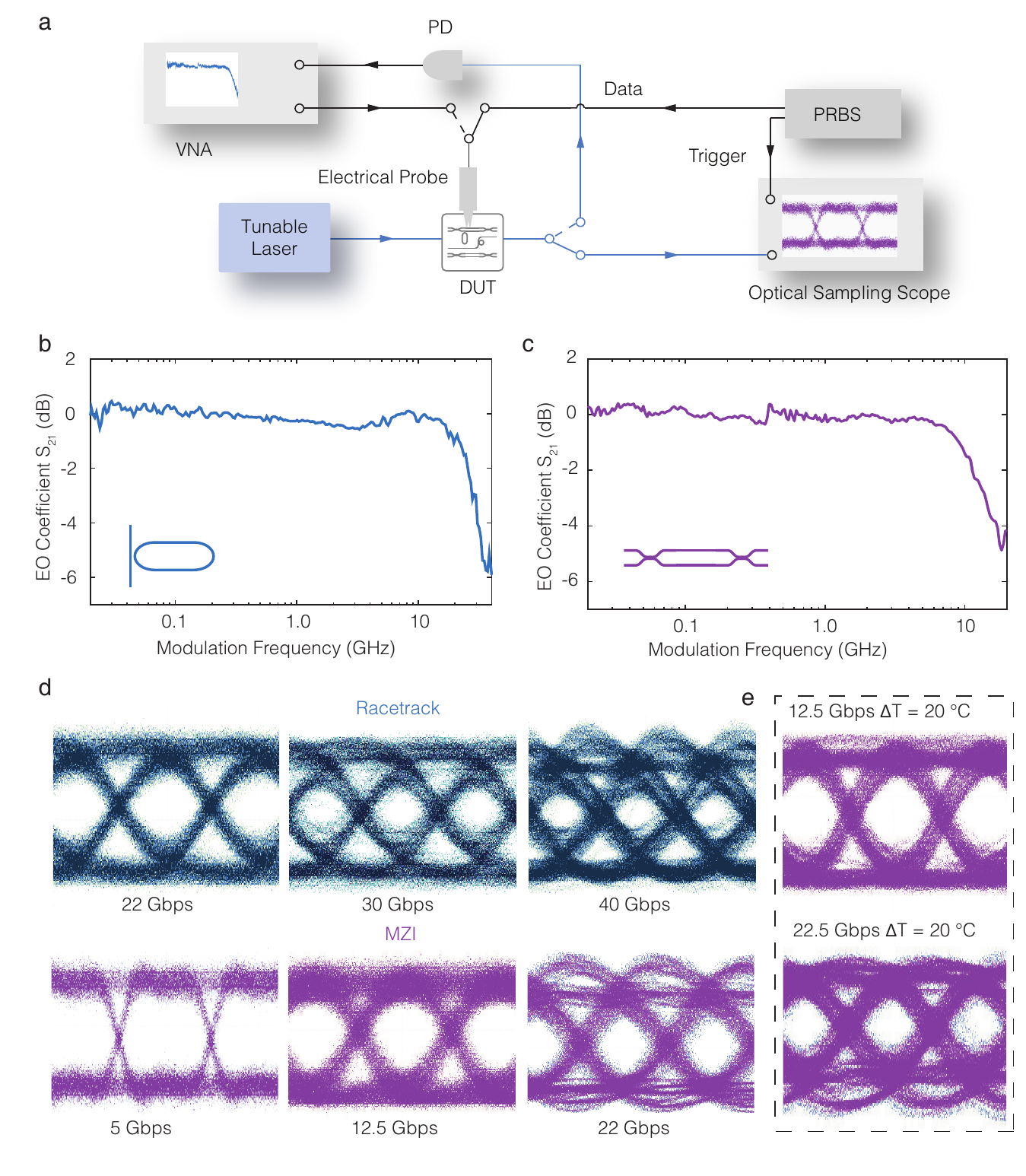}
  \caption{\label{fig4}\textbf{Bandwidth and high speed data operation} \textbf{a}. Simplified experimental setup for testing eye diagrams. Dashed lines indicate the signal path for electro-optic bandwidth measurement. \textbf{b-c}. Electro-optic bandwidths (S$_{21}$ parameter) of a race-track resonator with $Q$ $\sim$ 8,000 (a) and a 2 mm long MZI (b). The corresponding 3 dB bandwidths are 30 GHz and 15 GHz respectively. \textbf{d}. Eye diagrams of the racetrack (blue) and MZI (purple) modulator with data rates up to 40 Gbps and 22 Gbps. All eye diagrams are measured with 2$^7$-1 PRBS in a non-return-to-zero scheme with a 5.66 $V_{pp}$ electrical drive. \textbf{e}. Eye diagrams of the MZI modulator at 12.5 Gbps and 22 Gbps with the device heated up by 20 $\deg$C. The extinction ratios in d and e are 3 dB and 8 dB for racetrack resonator and MZI respectively. VNA: Vector network analyzer; PD: photodiode; PRBS: pseudo-random binary sequence; DUT: device under test.}
\end{figure*}

Our miniaturized LN devices exhibit high electro-optic bandwidths (S$_{21}$ parameter), which we characterize using a network analyzer and a high-speed photodiode (Fig. \ref{fig4}a and Methods). For a racetrack resonator modulator featuring a $Q$ factor of 8,000, we measure a 3 dB electro-optic bandwidth of 30 GHz (Fig. 4b). This value is limited by the cavity-photon lifetime of the resonator ($\sim$ 6 ps). We confirm the lifetime limited bandwidth by testing additional resonators with $Q$s of 5,700 and 18,000. The resulting 3 dB bandwidths are 40 GHz and 11 GHz respectively. The $Q$ factors are engineered from the intrinsic value by controlling the distance between the RF electrodes and the optical waveguide. The intrinsic RC bandwidth limit of the racetrack modulator is estimated to be over 100 GHz. For the 2 mm long MZI device with direct capacitive modulation, the measured electro-optic 3 dB bandwidth is $\sim$ 15 GHz (Fig. \ref{fig4}c). This is limited by the RC constant due to a larger capacitance ($\sim$ 0.2 pF) induced by the longer RF electrode used. Note that since our on-chip electrical resistance is rather small ($<$ 10 $\Omega$), the measured bandwidth is currently limited by the 50 $\Omega$ impedance of the network analyzer drive.

Our platform supports data transmission rates as high as 40 Gbps. Figure 4d displays non-return-to-zero (NRZ) open eye diagrams for both racetrack and MZI modulators at various data rates, obtained with $2^7-1$ (pseudo-) random binary sequence at 5.66 $V_{\textrm{pp}}$ (Methods). Because of the high signal quality, our devices can generally operate at data rates 1.5 times their 3 dB bandwidth, which translates to 40 Gbps and 22 Gbps for the racetrack and MZI devices respectively. The measured extinction ratios of our modulators are 3 dB and 8 dB with power consumptions (CV$^2$/4) of 240 fJ/bit and 1.6 pJ/bit respectively.

We confirm that our MZI modulator maintains the stable thermal properties of their bulk counterparts, due to the low thermo-optic coefficient of LN (3.9 $\times 10^{-5} K^{-1}$) \cite{Nikogosyan05}. We vary the temperature of our chip within a $\Delta T = 20~^{\circ}$C range, limited by the setup, and record the eye diagrams. We find that the MZI modulator is able to maintain an open eye diagram at the maximum data rate of 22 Gbps without any feedback to compensate for temperature drifts (open loop configuration) (Fig. \ref{fig4}e).

Our micrometer scale LN modulators feature high-bandwidth, excellent linearity, low voltage and good temperature stability. The monolithic LN photonic platform that we demonstrate here could lead to a paradigm shift in electro-optic modulator scalabilty, performance and design. For nearly a decade, existing LN modulator performance has been capped due to the non-ideal phase matching between the RF and optical fields \cite{Janner09}. The high dielectric constant of LN ($\epsilon_{\textrm{RF}} \sim 28$) dictates that RF fields in LN propagate much slower than optical fields ($\epsilon_{\textrm{opt}} \sim 4$) resulting in performane trade-off between bandwidth and driving voltage \cite{Janner09}. In our thin-film monolithic LN approach, instead, phase matching can be much better achieved since the electrical field primarily resides in the low dielectric SiO$_2$ ($\epsilon_{\textrm{opt}} \sim 4$) and readily propagates at nearly the same group velocity as light \cite{Janner09}. The thin-film micro-MZI modulators, with a phase-matched RF transmission line architecture, could simultaneously achieve ultra-high bandwidth ($>$ 60 GHz) and low modulation voltage ($\sim$ 1 V), and therefore are directly drivable with CMOS circuitry. The active micro-resonators and low loss waveguides could enable a chip-scale photonic circuit densely integrated with novel switches, filters, nonlinear wavelength sources and operates in a wide wavelength range (from visible to mid-IR). Furthermore, the ultra-compact footprint (as small as 30 $\mu$m $\times$ 30 $\mu$m) of micro-ring modulators is attractive for data center applications where real estate is at a premium. Our high-performance monolithic LN nanophotonic platform could become a practical cost-effective solution to meet the growing demands of next-generation data center, metro and long-haul optical telecommunications.

The authors thank J. Kahn for valuable discussions and H. Majedi for the use of VNA. This work is supported in part by the National Science Foundation (ECCS-1609549), Office of Naval Research MURI (N00014-15-1-2761) and Army Research Laboratory Center for Distributed Quantum Information (W911NF1520067). B.S. acknowledge support from the National Science Foundation - CIAN ERC (EEC-0812072). Device fabrication was performed at the center for nanoscale systems (CNS) at Harvard University.

\bibliography{reference}

\begin{thebibliography}{30}%
\makeatletter
\providecommand \@ifxundefined [1]{%
 \@ifx{#1\undefined}
}%
\providecommand \@ifnum [1]{%
 \ifnum #1\expandafter \@firstoftwo
 \else \expandafter \@secondoftwo
 \fi
}%
\providecommand \@ifx [1]{%
 \ifx #1\expandafter \@firstoftwo
 \else \expandafter \@secondoftwo
 \fi
}%
\providecommand \natexlab [1]{#1}%
\providecommand \enquote  [1]{``#1''}%
\providecommand \bibnamefont  [1]{#1}%
\providecommand \bibfnamefont [1]{#1}%
\providecommand \citenamefont [1]{#1}%
\providecommand \href@noop [0]{\@secondoftwo}%
\providecommand \href [0]{\begingroup \@sanitize@url \@href}%
\providecommand \@href[1]{\@@startlink{#1}\@@href}%
\providecommand \@@href[1]{\endgroup#1\@@endlink}%
\providecommand \@sanitize@url [0]{\catcode `\\12\catcode `\$12\catcode
  `\&12\catcode `\#12\catcode `\^12\catcode `\_12\catcode `\%12\relax}%
\providecommand \@@startlink[1]{}%
\providecommand \@@endlink[0]{}%
\providecommand \url  [0]{\begingroup\@sanitize@url \@url }%
\providecommand \@url [1]{\endgroup\@href {#1}{\urlprefix }}%
\providecommand \urlprefix  [0]{URL }%
\providecommand \Eprint [0]{\href }%
\providecommand \doibase [0]{http://dx.doi.org/}%
\providecommand \selectlanguage [0]{\@gobble}%
\providecommand \bibinfo  [0]{\@secondoftwo}%
\providecommand \bibfield  [0]{\@secondoftwo}%
\providecommand \translation [1]{[#1]}%
\providecommand \BibitemOpen [0]{}%
\providecommand \bibitemStop [0]{}%
\providecommand \bibitemNoStop [0]{.\EOS\space}%
\providecommand \EOS [0]{\spacefactor3000\relax}%
\providecommand \BibitemShut  [1]{\csname bibitem#1\endcsname}%
\let\auto@bib@innerbib\@empty
\bibitem [{\citenamefont {Wooten}\ \emph {et~al.}(2000)\citenamefont {Wooten},
  \citenamefont {Kissa}, \citenamefont {Yi-Yan}, \citenamefont {Murphy},
  \citenamefont {Lafaw}, \citenamefont {Hallemeier}, \citenamefont {Maack},
  \citenamefont {Attanasio}, \citenamefont {Fritz}, \citenamefont {McBrien},\
  and\ \citenamefont {Bossi}}]{Wooten00}%
  \BibitemOpen
  \bibfield  {author} {\bibinfo {author} {\bibfnamefont {E.~L.}\ \bibnamefont
  {Wooten}}, \bibinfo {author} {\bibfnamefont {K.~M.}\ \bibnamefont {Kissa}},
  \bibinfo {author} {\bibfnamefont {A.}~\bibnamefont {Yi-Yan}}, \bibinfo
  {author} {\bibfnamefont {E.~J.}\ \bibnamefont {Murphy}}, \bibinfo {author}
  {\bibfnamefont {D.~A.}\ \bibnamefont {Lafaw}}, \bibinfo {author}
  {\bibfnamefont {P.~F.}\ \bibnamefont {Hallemeier}}, \bibinfo {author}
  {\bibfnamefont {D.}~\bibnamefont {Maack}}, \bibinfo {author} {\bibfnamefont
  {D.~V.}\ \bibnamefont {Attanasio}}, \bibinfo {author} {\bibfnamefont {D.~J.}\
  \bibnamefont {Fritz}}, \bibinfo {author} {\bibfnamefont {G.~J.}\ \bibnamefont
  {McBrien}}, \ and\ \bibinfo {author} {\bibfnamefont {D.~E.}\ \bibnamefont
  {Bossi}},\ }\href {\doibase 10.1109/2944.826874} {\bibfield  {journal}
  {\bibinfo  {journal} {Selected Topics in Quantum Electronics, IEEE Journal
  of}\ }\textbf {\bibinfo {volume} {6}},\ \bibinfo {pages} {69} (\bibinfo
  {year} {2000})}\BibitemShut {NoStop}%
\bibitem [{\citenamefont {Janner}\ \emph {et~al.}(2009)\citenamefont {Janner},
  \citenamefont {Tulli}, \citenamefont {García-Granda}, \citenamefont
  {Belmonte},\ and\ \citenamefont {Pruneri}}]{Janner09}%
  \BibitemOpen
  \bibfield  {author} {\bibinfo {author} {\bibfnamefont {D.}~\bibnamefont
  {Janner}}, \bibinfo {author} {\bibfnamefont {D.}~\bibnamefont {Tulli}},
  \bibinfo {author} {\bibfnamefont {M.}~\bibnamefont {García-Granda}},
  \bibinfo {author} {\bibfnamefont {M.}~\bibnamefont {Belmonte}}, \ and\
  \bibinfo {author} {\bibfnamefont {V.}~\bibnamefont {Pruneri}},\ }\href
  {\doibase 10.1002/lpor.200810073} {\bibfield  {journal} {\bibinfo  {journal}
  {Laser \& Photonics Reviews}\ }\textbf {\bibinfo {volume} {3}},\ \bibinfo
  {pages} {301} (\bibinfo {year} {2009})}\BibitemShut {NoStop}%
\bibitem [{\citenamefont {Nikogosyan}(2005)}]{Nikogosyan05}%
  \BibitemOpen
  \bibfield  {author} {\bibinfo {author} {\bibfnamefont {D.~N.}\ \bibnamefont
  {Nikogosyan}},\ }\href@noop {} {\emph {\bibinfo {title} {Nonlinear Optical
  Crystals: A Complete Survey}}}\ (\bibinfo  {publisher} {Springer-Science},\
  \bibinfo {address} {New York},\ \bibinfo {year} {2005})\BibitemShut {NoStop}%
\bibitem [{\citenamefont {Xu}\ \emph {et~al.}(2005)\citenamefont {Xu},
  \citenamefont {Schmidt}, \citenamefont {Pradhan},\ and\ \citenamefont
  {Lipson}}]{Xu05}%
  \BibitemOpen
  \bibfield  {author} {\bibinfo {author} {\bibfnamefont {Q.}~\bibnamefont
  {Xu}}, \bibinfo {author} {\bibfnamefont {B.}~\bibnamefont {Schmidt}},
  \bibinfo {author} {\bibfnamefont {S.}~\bibnamefont {Pradhan}}, \ and\
  \bibinfo {author} {\bibfnamefont {M.}~\bibnamefont {Lipson}},\ }\href@noop {}
  {\bibfield  {journal} {\bibinfo  {journal} {Nature}\ }\textbf {\bibinfo
  {volume} {435}},\ \bibinfo {pages} {325} (\bibinfo {year}
  {2005})}\BibitemShut {NoStop}%
\bibitem [{\citenamefont {Timurdogan}\ \emph {et~al.}(2014)\citenamefont
  {Timurdogan}, \citenamefont {Sorace-Agaskar}, \citenamefont {Sun},
  \citenamefont {Shah~Hosseini}, \citenamefont {Biberman},\ and\ \citenamefont
  {Watts}}]{Timurdogan14}%
  \BibitemOpen
  \bibfield  {author} {\bibinfo {author} {\bibfnamefont {E.}~\bibnamefont
  {Timurdogan}}, \bibinfo {author} {\bibfnamefont {C.~M.}\ \bibnamefont
  {Sorace-Agaskar}}, \bibinfo {author} {\bibfnamefont {J.}~\bibnamefont {Sun}},
  \bibinfo {author} {\bibfnamefont {E.}~\bibnamefont {Shah~Hosseini}}, \bibinfo
  {author} {\bibfnamefont {A.}~\bibnamefont {Biberman}}, \ and\ \bibinfo
  {author} {\bibfnamefont {M.~R.}\ \bibnamefont {Watts}},\ }\href {\doibase
  10.1038/ncomms5008} {\bibfield  {journal} {\bibinfo  {journal} {Nature
  Communications}\ }\textbf {\bibinfo {volume} {5}},\ \bibinfo {pages} {4008}
  (\bibinfo {year} {2014})}\BibitemShut {NoStop}%
\bibitem [{\citenamefont {Zhang}\ \emph {et~al.}(2016)\citenamefont {Zhang},
  \citenamefont {Morton}, \citenamefont {Khurgin}, \citenamefont {Peters},\
  and\ \citenamefont {Bowers}}]{Zhang16}%
  \BibitemOpen
  \bibfield  {author} {\bibinfo {author} {\bibfnamefont {C.}~\bibnamefont
  {Zhang}}, \bibinfo {author} {\bibfnamefont {P.~A.}\ \bibnamefont {Morton}},
  \bibinfo {author} {\bibfnamefont {J.~B.}\ \bibnamefont {Khurgin}}, \bibinfo
  {author} {\bibfnamefont {J.~D.}\ \bibnamefont {Peters}}, \ and\ \bibinfo
  {author} {\bibfnamefont {J.~E.}\ \bibnamefont {Bowers}},\ }\href {\doibase
  10.1364/OPTICA.3.001483} {\bibfield  {journal} {\bibinfo  {journal} {Optica}\
  }\textbf {\bibinfo {volume} {3}},\ \bibinfo {pages} {1483} (\bibinfo {year}
  {2016})}\BibitemShut {NoStop}%
\bibitem [{\citenamefont {Sun}\ \emph {et~al.}(2015)\citenamefont {Sun},
  \citenamefont {Wade}, \citenamefont {Lee}, \citenamefont {Orcutt},
  \citenamefont {Alloatti}, \citenamefont {Georgas}, \citenamefont {Waterman},
  \citenamefont {Shainline}, \citenamefont {Avizienis}, \citenamefont {Lin},
  \citenamefont {Moss}, \citenamefont {Kumar}, \citenamefont {Pavanello},
  \citenamefont {Atabaki}, \citenamefont {Cook}, \citenamefont {Ou},
  \citenamefont {Leu}, \citenamefont {Chen}, \citenamefont {Asanović},
  \citenamefont {Ram}, \citenamefont {Popović},\ and\ \citenamefont
  {Stojanović}}]{Sun15}%
  \BibitemOpen
  \bibfield  {author} {\bibinfo {author} {\bibfnamefont {C.}~\bibnamefont
  {Sun}}, \bibinfo {author} {\bibfnamefont {M.~T.}\ \bibnamefont {Wade}},
  \bibinfo {author} {\bibfnamefont {Y.}~\bibnamefont {Lee}}, \bibinfo {author}
  {\bibfnamefont {J.~S.}\ \bibnamefont {Orcutt}}, \bibinfo {author}
  {\bibfnamefont {L.}~\bibnamefont {Alloatti}}, \bibinfo {author}
  {\bibfnamefont {M.~S.}\ \bibnamefont {Georgas}}, \bibinfo {author}
  {\bibfnamefont {A.~S.}\ \bibnamefont {Waterman}}, \bibinfo {author}
  {\bibfnamefont {J.~M.}\ \bibnamefont {Shainline}}, \bibinfo {author}
  {\bibfnamefont {R.~R.}\ \bibnamefont {Avizienis}}, \bibinfo {author}
  {\bibfnamefont {S.}~\bibnamefont {Lin}}, \bibinfo {author} {\bibfnamefont
  {B.~R.}\ \bibnamefont {Moss}}, \bibinfo {author} {\bibfnamefont
  {R.}~\bibnamefont {Kumar}}, \bibinfo {author} {\bibfnamefont
  {F.}~\bibnamefont {Pavanello}}, \bibinfo {author} {\bibfnamefont {A.~H.}\
  \bibnamefont {Atabaki}}, \bibinfo {author} {\bibfnamefont {H.~M.}\
  \bibnamefont {Cook}}, \bibinfo {author} {\bibfnamefont {A.~J.}\ \bibnamefont
  {Ou}}, \bibinfo {author} {\bibfnamefont {J.~C.}\ \bibnamefont {Leu}},
  \bibinfo {author} {\bibfnamefont {Y.-H.}\ \bibnamefont {Chen}}, \bibinfo
  {author} {\bibfnamefont {K.}~\bibnamefont {Asanović}}, \bibinfo {author}
  {\bibfnamefont {R.~J.}\ \bibnamefont {Ram}}, \bibinfo {author} {\bibfnamefont
  {M.~A.}\ \bibnamefont {Popović}}, \ and\ \bibinfo {author} {\bibfnamefont
  {V.~M.}\ \bibnamefont {Stojanović}},\ }\href {\doibase 10.1038/nature16454
  http://www.nature.com/nature/journal/v528/n7583/abs/nature16454.html#supplementary-information}
  {\bibfield  {journal} {\bibinfo  {journal} {Nature}\ }\textbf {\bibinfo
  {volume} {528}},\ \bibinfo {pages} {534} (\bibinfo {year}
  {2015})}\BibitemShut {NoStop}%
\bibitem [{\citenamefont {Rolland}\ \emph {et~al.}(1993)\citenamefont
  {Rolland}, \citenamefont {Moore}, \citenamefont {Shepherd},\ and\
  \citenamefont {Hillier}}]{Rolland93}%
  \BibitemOpen
  \bibfield  {author} {\bibinfo {author} {\bibfnamefont {C.}~\bibnamefont
  {Rolland}}, \bibinfo {author} {\bibfnamefont {R.~S.}\ \bibnamefont {Moore}},
  \bibinfo {author} {\bibfnamefont {F.}~\bibnamefont {Shepherd}}, \ and\
  \bibinfo {author} {\bibfnamefont {G.}~\bibnamefont {Hillier}},\ }\href
  {\doibase 10.1049/el:19930315} {\bibfield  {journal} {\bibinfo  {journal}
  {Electronics Letters}\ }\textbf {\bibinfo {volume} {29}},\ \bibinfo {pages}
  {471} (\bibinfo {year} {1993})}\BibitemShut {NoStop}%
\bibitem [{\citenamefont {Aoki}\ \emph {et~al.}(1993)\citenamefont {Aoki},
  \citenamefont {Suzuki}, \citenamefont {Sano}, \citenamefont {Kawano},
  \citenamefont {Ido}, \citenamefont {Taniwatari}, \citenamefont {Uomi},\ and\
  \citenamefont {Takai}}]{Aoki93}%
  \BibitemOpen
  \bibfield  {author} {\bibinfo {author} {\bibfnamefont {M.}~\bibnamefont
  {Aoki}}, \bibinfo {author} {\bibfnamefont {M.}~\bibnamefont {Suzuki}},
  \bibinfo {author} {\bibfnamefont {H.}~\bibnamefont {Sano}}, \bibinfo {author}
  {\bibfnamefont {T.}~\bibnamefont {Kawano}}, \bibinfo {author} {\bibfnamefont
  {T.}~\bibnamefont {Ido}}, \bibinfo {author} {\bibfnamefont {T.}~\bibnamefont
  {Taniwatari}}, \bibinfo {author} {\bibfnamefont {K.}~\bibnamefont {Uomi}}, \
  and\ \bibinfo {author} {\bibfnamefont {A.}~\bibnamefont {Takai}},\ }\href
  {\doibase 10.1109/3.234473} {\bibfield  {journal} {\bibinfo  {journal} {IEEE
  Journal of Quantum Electronics}\ }\textbf {\bibinfo {volume} {29}},\ \bibinfo
  {pages} {2088} (\bibinfo {year} {1993})}\BibitemShut {NoStop}%
\bibitem [{\citenamefont {Kikuchi}\ \emph {et~al.}(2012)\citenamefont
  {Kikuchi}, \citenamefont {Yamada}, \citenamefont {Shibata},\ and\
  \citenamefont {Ishii}}]{Kikuchi12}%
  \BibitemOpen
  \bibfield  {author} {\bibinfo {author} {\bibfnamefont {N.}~\bibnamefont
  {Kikuchi}}, \bibinfo {author} {\bibfnamefont {E.}~\bibnamefont {Yamada}},
  \bibinfo {author} {\bibfnamefont {Y.}~\bibnamefont {Shibata}}, \ and\
  \bibinfo {author} {\bibfnamefont {H.}~\bibnamefont {Ishii}},\ }in\ \href
  {\doibase 10.1109/CSICS.2012.6340090} {\emph {\bibinfo {booktitle} {2012 IEEE
  Compound Semiconductor Integrated Circuit Symposium (CSICS)}}}\ (\bibinfo
  {year} {2012})\ pp.\ \bibinfo {pages} {1--4}\BibitemShut {NoStop}%
\bibitem [{\citenamefont {Xiong}\ \emph {et~al.}(2012)\citenamefont {Xiong},
  \citenamefont {Pernice},\ and\ \citenamefont {Tang}}]{Xiong12}%
  \BibitemOpen
  \bibfield  {author} {\bibinfo {author} {\bibfnamefont {C.}~\bibnamefont
  {Xiong}}, \bibinfo {author} {\bibfnamefont {W.~H.~P.}\ \bibnamefont
  {Pernice}}, \ and\ \bibinfo {author} {\bibfnamefont {H.~X.}\ \bibnamefont
  {Tang}},\ }\href {\doibase 10.1021/nl3011885} {\bibfield  {journal} {\bibinfo
   {journal} {Nano Letters}\ }\textbf {\bibinfo {volume} {12}},\ \bibinfo
  {pages} {3562} (\bibinfo {year} {2012})}\BibitemShut {NoStop}%
\bibitem [{\citenamefont {HaffnerC}\ \emph {et~al.}(2015)\citenamefont
  {HaffnerC}, \citenamefont {HeniW}, \citenamefont {FedoryshynY}, \citenamefont
  {NiegemannJ}, \citenamefont {MelikyanA}, \citenamefont {Elder}, \citenamefont
  {BaeuerleB}, \citenamefont {SalaminY}, \citenamefont {JostenA}, \citenamefont
  {KochU}, \citenamefont {HoessbacherC}, \citenamefont {DucryF}, \citenamefont
  {JuchliL}, \citenamefont {EmborasA}, \citenamefont {HillerkussD},
  \citenamefont {KohlM}, \citenamefont {Dalton}, \citenamefont {HafnerC},\ and\
  \citenamefont {LeutholdJ}}]{Haffner15}%
  \BibitemOpen
  \bibfield  {author} {\bibinfo {author} {\bibnamefont {HaffnerC}}, \bibinfo
  {author} {\bibnamefont {HeniW}}, \bibinfo {author} {\bibnamefont
  {FedoryshynY}}, \bibinfo {author} {\bibnamefont {NiegemannJ}}, \bibinfo
  {author} {\bibnamefont {MelikyanA}}, \bibinfo {author} {\bibfnamefont
  {D.~L.}\ \bibnamefont {Elder}}, \bibinfo {author} {\bibnamefont {BaeuerleB}},
  \bibinfo {author} {\bibnamefont {SalaminY}}, \bibinfo {author} {\bibnamefont
  {JostenA}}, \bibinfo {author} {\bibnamefont {KochU}}, \bibinfo {author}
  {\bibnamefont {HoessbacherC}}, \bibinfo {author} {\bibnamefont {DucryF}},
  \bibinfo {author} {\bibnamefont {JuchliL}}, \bibinfo {author} {\bibnamefont
  {EmborasA}}, \bibinfo {author} {\bibnamefont {HillerkussD}}, \bibinfo
  {author} {\bibnamefont {KohlM}}, \bibinfo {author} {\bibfnamefont {L.~R.}\
  \bibnamefont {Dalton}}, \bibinfo {author} {\bibnamefont {HafnerC}}, \ and\
  \bibinfo {author} {\bibnamefont {LeutholdJ}},\ }\href {\doibase
  10.1038/nphoton.2015.127
  http://www.nature.com/nphoton/journal/v9/n8/abs/nphoton.2015.127.html#supplementary-information}
  {\bibfield  {journal} {\bibinfo  {journal} {Nat Photon}\ }\textbf {\bibinfo
  {volume} {9}},\ \bibinfo {pages} {525} (\bibinfo {year} {2015})}\BibitemShut
  {NoStop}%
\bibitem [{\citenamefont {Liu}\ \emph {et~al.}(2011)\citenamefont {Liu},
  \citenamefont {Yin}, \citenamefont {Ulin-Avila}, \citenamefont {Geng},
  \citenamefont {Zentgraf}, \citenamefont {Ju}, \citenamefont {Wang},\ and\
  \citenamefont {Zhang}}]{Liu11}%
  \BibitemOpen
  \bibfield  {author} {\bibinfo {author} {\bibfnamefont {M.}~\bibnamefont
  {Liu}}, \bibinfo {author} {\bibfnamefont {X.}~\bibnamefont {Yin}}, \bibinfo
  {author} {\bibfnamefont {E.}~\bibnamefont {Ulin-Avila}}, \bibinfo {author}
  {\bibfnamefont {B.}~\bibnamefont {Geng}}, \bibinfo {author} {\bibfnamefont
  {T.}~\bibnamefont {Zentgraf}}, \bibinfo {author} {\bibfnamefont
  {L.}~\bibnamefont {Ju}}, \bibinfo {author} {\bibfnamefont {F.}~\bibnamefont
  {Wang}}, \ and\ \bibinfo {author} {\bibfnamefont {X.}~\bibnamefont {Zhang}},\
  }\href@noop {} {\bibfield  {journal} {\bibinfo  {journal} {Nature}\ }\textbf
  {\bibinfo {volume} {474}},\ \bibinfo {pages} {64} (\bibinfo {year}
  {2011})}\BibitemShut {NoStop}%
\bibitem [{\citenamefont {Phare}\ \emph {et~al.}(2015)\citenamefont {Phare},
  \citenamefont {Daniel~Lee}, \citenamefont {Cardenas},\ and\ \citenamefont
  {Lipson}}]{Phare15}%
  \BibitemOpen
  \bibfield  {author} {\bibinfo {author} {\bibfnamefont {C.~T.}\ \bibnamefont
  {Phare}}, \bibinfo {author} {\bibfnamefont {Y.-H.}\ \bibnamefont
  {Daniel~Lee}}, \bibinfo {author} {\bibfnamefont {J.}~\bibnamefont
  {Cardenas}}, \ and\ \bibinfo {author} {\bibfnamefont {M.}~\bibnamefont
  {Lipson}},\ }\href {\doibase 10.1038/nphoton.2015.122
  http://www.nature.com/nphoton/journal/v9/n8/abs/nphoton.2015.122.html#supplementary-information}
  {\bibfield  {journal} {\bibinfo  {journal} {Nat Photon}\ }\textbf {\bibinfo
  {volume} {9}},\ \bibinfo {pages} {511} (\bibinfo {year} {2015})}\BibitemShut
  {NoStop}%
\bibitem [{\citenamefont {Novoselov}\ \emph {et~al.}(2012)\citenamefont
  {Novoselov}, \citenamefont {Falko}, \citenamefont {Colombo}, \citenamefont
  {Gellert}, \citenamefont {Schwab},\ and\ \citenamefont {Kim}}]{Novoselov12}%
  \BibitemOpen
  \bibfield  {author} {\bibinfo {author} {\bibfnamefont {K.~S.}\ \bibnamefont
  {Novoselov}}, \bibinfo {author} {\bibfnamefont {V.~I.}\ \bibnamefont
  {Falko}}, \bibinfo {author} {\bibfnamefont {L.}~\bibnamefont {Colombo}},
  \bibinfo {author} {\bibfnamefont {P.~R.}\ \bibnamefont {Gellert}}, \bibinfo
  {author} {\bibfnamefont {M.~G.}\ \bibnamefont {Schwab}}, \ and\ \bibinfo
  {author} {\bibfnamefont {K.}~\bibnamefont {Kim}},\ }\href@noop {} {\bibfield
  {journal} {\bibinfo  {journal} {Nature}\ }\textbf {\bibinfo {volume} {490}},\
  \bibinfo {pages} {192} (\bibinfo {year} {2012})}\BibitemShut {NoStop}%
\bibitem [{\citenamefont {Lee}\ \emph {et~al.}(2002)\citenamefont {Lee},
  \citenamefont {Katz}, \citenamefont {Erben}, \citenamefont {Gill},
  \citenamefont {Gopalan}, \citenamefont {Heber},\ and\ \citenamefont
  {McGee}}]{Lee02}%
  \BibitemOpen
  \bibfield  {author} {\bibinfo {author} {\bibfnamefont {M.}~\bibnamefont
  {Lee}}, \bibinfo {author} {\bibfnamefont {H.~E.}\ \bibnamefont {Katz}},
  \bibinfo {author} {\bibfnamefont {C.}~\bibnamefont {Erben}}, \bibinfo
  {author} {\bibfnamefont {D.~M.}\ \bibnamefont {Gill}}, \bibinfo {author}
  {\bibfnamefont {P.}~\bibnamefont {Gopalan}}, \bibinfo {author} {\bibfnamefont
  {J.~D.}\ \bibnamefont {Heber}}, \ and\ \bibinfo {author} {\bibfnamefont
  {D.~J.}\ \bibnamefont {McGee}},\ }\href@noop {} {\bibfield  {journal}
  {\bibinfo  {journal} {Science}\ }\textbf {\bibinfo {volume} {298}},\ \bibinfo
  {pages} {1401} (\bibinfo {year} {2002})}\BibitemShut {NoStop}%
\bibitem [{\citenamefont {Clark}\ and\ \citenamefont
  {Lanzani}(2010)}]{Clark10}%
  \BibitemOpen
  \bibfield  {author} {\bibinfo {author} {\bibfnamefont {J.}~\bibnamefont
  {Clark}}\ and\ \bibinfo {author} {\bibfnamefont {G.}~\bibnamefont
  {Lanzani}},\ }\href@noop {} {\bibfield  {journal} {\bibinfo  {journal} {Nat
  Photon}\ }\textbf {\bibinfo {volume} {4}},\ \bibinfo {pages} {438} (\bibinfo
  {year} {2010})}\BibitemShut {NoStop}%
\bibitem [{\citenamefont {Schmidt}\ and\ \citenamefont
  {Kaminow}(1974)}]{Schmidt74}%
  \BibitemOpen
  \bibfield  {author} {\bibinfo {author} {\bibfnamefont {R.~V.}\ \bibnamefont
  {Schmidt}}\ and\ \bibinfo {author} {\bibfnamefont {I.~P.}\ \bibnamefont
  {Kaminow}},\ }\href {\doibase doi:http://dx.doi.org/10.1063/1.1655547}
  {\bibfield  {journal} {\bibinfo  {journal} {Applied Physics Letters}\
  }\textbf {\bibinfo {volume} {25}},\ \bibinfo {pages} {458} (\bibinfo {year}
  {1974})}\BibitemShut {NoStop}%
\bibitem [{\citenamefont {Poberaj}\ \emph {et~al.}(2012)\citenamefont
  {Poberaj}, \citenamefont {Hu}, \citenamefont {Sohler},\ and\ \citenamefont
  {Günter}}]{Poberaj12}%
  \BibitemOpen
  \bibfield  {author} {\bibinfo {author} {\bibfnamefont {G.}~\bibnamefont
  {Poberaj}}, \bibinfo {author} {\bibfnamefont {H.}~\bibnamefont {Hu}},
  \bibinfo {author} {\bibfnamefont {W.}~\bibnamefont {Sohler}}, \ and\ \bibinfo
  {author} {\bibfnamefont {P.}~\bibnamefont {Günter}},\ }\href {\doibase
  10.1002/lpor.201100035} {\bibfield  {journal} {\bibinfo  {journal} {Laser \&
  Photonics Reviews}\ }\textbf {\bibinfo {volume} {6}},\ \bibinfo {pages} {488}
  (\bibinfo {year} {2012})}\BibitemShut {NoStop}%
\bibitem [{\citenamefont {Guarino}\ \emph {et~al.}(2007)\citenamefont
  {Guarino}, \citenamefont {Poberaj}, \citenamefont {Rezzonico}, \citenamefont
  {Degl'Innocenti},\ and\ \citenamefont {Gunter}}]{Guarino07}%
  \BibitemOpen
  \bibfield  {author} {\bibinfo {author} {\bibfnamefont {A.}~\bibnamefont
  {Guarino}}, \bibinfo {author} {\bibfnamefont {G.}~\bibnamefont {Poberaj}},
  \bibinfo {author} {\bibfnamefont {D.}~\bibnamefont {Rezzonico}}, \bibinfo
  {author} {\bibfnamefont {R.}~\bibnamefont {Degl'Innocenti}}, \ and\ \bibinfo
  {author} {\bibfnamefont {P.}~\bibnamefont {Gunter}},\ }\href {\doibase
  http://www.nature.com/nphoton/journal/v1/n7/suppinfo/nphoton.2007.93_S1.html}
  {\bibfield  {journal} {\bibinfo  {journal} {Nat Photon}\ }\textbf {\bibinfo
  {volume} {1}},\ \bibinfo {pages} {407} (\bibinfo {year} {2007})}\BibitemShut
  {NoStop}%
\bibitem [{\citenamefont {Chen}\ \emph {et~al.}(2014)\citenamefont {Chen},
  \citenamefont {Xu}, \citenamefont {Wood},\ and\ \citenamefont
  {Reano}}]{Chen14}%
  \BibitemOpen
  \bibfield  {author} {\bibinfo {author} {\bibfnamefont {L.}~\bibnamefont
  {Chen}}, \bibinfo {author} {\bibfnamefont {Q.}~\bibnamefont {Xu}}, \bibinfo
  {author} {\bibfnamefont {M.~G.}\ \bibnamefont {Wood}}, \ and\ \bibinfo
  {author} {\bibfnamefont {R.~M.}\ \bibnamefont {Reano}},\ }\href {\doibase
  10.1364/OPTICA.1.000112} {\bibfield  {journal} {\bibinfo  {journal} {Optica}\
  }\textbf {\bibinfo {volume} {1}},\ \bibinfo {pages} {112} (\bibinfo {year}
  {2014})}\BibitemShut {NoStop}%
\bibitem [{\citenamefont {Rao}\ \emph {et~al.}(2015)\citenamefont {Rao},
  \citenamefont {Patil}, \citenamefont {Chiles}, \citenamefont {Malinowski},
  \citenamefont {Novak}, \citenamefont {Richardson}, \citenamefont {Rabiei},\
  and\ \citenamefont {Fathpour}}]{Rao15}%
  \BibitemOpen
  \bibfield  {author} {\bibinfo {author} {\bibfnamefont {A.}~\bibnamefont
  {Rao}}, \bibinfo {author} {\bibfnamefont {A.}~\bibnamefont {Patil}}, \bibinfo
  {author} {\bibfnamefont {J.}~\bibnamefont {Chiles}}, \bibinfo {author}
  {\bibfnamefont {M.}~\bibnamefont {Malinowski}}, \bibinfo {author}
  {\bibfnamefont {S.}~\bibnamefont {Novak}}, \bibinfo {author} {\bibfnamefont
  {K.}~\bibnamefont {Richardson}}, \bibinfo {author} {\bibfnamefont
  {P.}~\bibnamefont {Rabiei}}, \ and\ \bibinfo {author} {\bibfnamefont
  {S.}~\bibnamefont {Fathpour}},\ }\href {\doibase 10.1364/OE.23.022746}
  {\bibfield  {journal} {\bibinfo  {journal} {Optics Express}\ }\textbf
  {\bibinfo {volume} {23}},\ \bibinfo {pages} {22746} (\bibinfo {year}
  {2015})}\BibitemShut {NoStop}%
\bibitem [{\citenamefont {Jin}\ \emph {et~al.}(2016)\citenamefont {Jin},
  \citenamefont {Xu}, \citenamefont {Zhang},\ and\ \citenamefont {Li}}]{Jin16}%
  \BibitemOpen
  \bibfield  {author} {\bibinfo {author} {\bibfnamefont {S.}~\bibnamefont
  {Jin}}, \bibinfo {author} {\bibfnamefont {L.}~\bibnamefont {Xu}}, \bibinfo
  {author} {\bibfnamefont {H.}~\bibnamefont {Zhang}}, \ and\ \bibinfo {author}
  {\bibfnamefont {Y.}~\bibnamefont {Li}},\ }\href {\doibase
  10.1109/LPT.2015.2507136} {\bibfield  {journal} {\bibinfo  {journal} {IEEE
  Photonics Technology Letters}\ }\textbf {\bibinfo {volume} {28}},\ \bibinfo
  {pages} {736} (\bibinfo {year} {2016})}\BibitemShut {NoStop}%
\bibitem [{\citenamefont {Rao}\ \emph {et~al.}(2016)\citenamefont {Rao},
  \citenamefont {Patil}, \citenamefont {Rabiei}, \citenamefont {Honardoost},
  \citenamefont {DeSalvo}, \citenamefont {Paolella},\ and\ \citenamefont
  {Fathpour}}]{Rao16}%
  \BibitemOpen
  \bibfield  {author} {\bibinfo {author} {\bibfnamefont {A.}~\bibnamefont
  {Rao}}, \bibinfo {author} {\bibfnamefont {A.}~\bibnamefont {Patil}}, \bibinfo
  {author} {\bibfnamefont {P.}~\bibnamefont {Rabiei}}, \bibinfo {author}
  {\bibfnamefont {A.}~\bibnamefont {Honardoost}}, \bibinfo {author}
  {\bibfnamefont {R.}~\bibnamefont {DeSalvo}}, \bibinfo {author} {\bibfnamefont
  {A.}~\bibnamefont {Paolella}}, \ and\ \bibinfo {author} {\bibfnamefont
  {S.}~\bibnamefont {Fathpour}},\ }\href {\doibase 10.1364/OL.41.005700}
  {\bibfield  {journal} {\bibinfo  {journal} {Optics Letters}\ }\textbf
  {\bibinfo {volume} {41}},\ \bibinfo {pages} {5700} (\bibinfo {year}
  {2016})}\BibitemShut {NoStop}%
\bibitem [{\citenamefont {Wang}\ \emph {et~al.}(2015)\citenamefont {Wang},
  \citenamefont {Bo}, \citenamefont {Wan}, \citenamefont {Li}, \citenamefont
  {Gao}, \citenamefont {Li}, \citenamefont {Zhang},\ and\ \citenamefont
  {Xu}}]{Wang15}%
  \BibitemOpen
  \bibfield  {author} {\bibinfo {author} {\bibfnamefont {J.}~\bibnamefont
  {Wang}}, \bibinfo {author} {\bibfnamefont {F.}~\bibnamefont {Bo}}, \bibinfo
  {author} {\bibfnamefont {S.}~\bibnamefont {Wan}}, \bibinfo {author}
  {\bibfnamefont {W.}~\bibnamefont {Li}}, \bibinfo {author} {\bibfnamefont
  {F.}~\bibnamefont {Gao}}, \bibinfo {author} {\bibfnamefont {J.}~\bibnamefont
  {Li}}, \bibinfo {author} {\bibfnamefont {G.}~\bibnamefont {Zhang}}, \ and\
  \bibinfo {author} {\bibfnamefont {J.}~\bibnamefont {Xu}},\ }\href {\doibase
  10.1364/OE.23.023072} {\bibfield  {journal} {\bibinfo  {journal} {Optics
  Express}\ }\textbf {\bibinfo {volume} {23}},\ \bibinfo {pages} {23072}
  (\bibinfo {year} {2015})}\BibitemShut {NoStop}%
\bibitem [{\citenamefont {Cai}\ \emph {et~al.}(2016)\citenamefont {Cai},
  \citenamefont {Kang},\ and\ \citenamefont {Hu}}]{Cai16}%
  \BibitemOpen
  \bibfield  {author} {\bibinfo {author} {\bibfnamefont {L.}~\bibnamefont
  {Cai}}, \bibinfo {author} {\bibfnamefont {Y.}~\bibnamefont {Kang}}, \ and\
  \bibinfo {author} {\bibfnamefont {H.}~\bibnamefont {Hu}},\ }\href {\doibase
  10.1364/OE.24.004640} {\bibfield  {journal} {\bibinfo  {journal} {Optics
  Express}\ }\textbf {\bibinfo {volume} {24}},\ \bibinfo {pages} {4640}
  (\bibinfo {year} {2016})}\BibitemShut {NoStop}%
\bibitem [{\citenamefont {Wang}\ \emph {et~al.}(2014)\citenamefont {Wang},
  \citenamefont {Burek}, \citenamefont {Lin}, \citenamefont {Atikian},
  \citenamefont {Venkataraman}, \citenamefont {Huang}, \citenamefont {Stark},\
  and\ \citenamefont {Lončar}}]{Wang14}%
  \BibitemOpen
  \bibfield  {author} {\bibinfo {author} {\bibfnamefont {C.}~\bibnamefont
  {Wang}}, \bibinfo {author} {\bibfnamefont {M.~J.}\ \bibnamefont {Burek}},
  \bibinfo {author} {\bibfnamefont {Z.}~\bibnamefont {Lin}}, \bibinfo {author}
  {\bibfnamefont {H.~A.}\ \bibnamefont {Atikian}}, \bibinfo {author}
  {\bibfnamefont {V.}~\bibnamefont {Venkataraman}}, \bibinfo {author}
  {\bibfnamefont {I.~C.}\ \bibnamefont {Huang}}, \bibinfo {author}
  {\bibfnamefont {P.}~\bibnamefont {Stark}}, \ and\ \bibinfo {author}
  {\bibfnamefont {M.}~\bibnamefont {Lončar}},\ }\href {\doibase
  10.1364/OE.22.030924} {\bibfield  {journal} {\bibinfo  {journal} {Optics
  Express}\ }\textbf {\bibinfo {volume} {22}},\ \bibinfo {pages} {30924}
  (\bibinfo {year} {2014})}\BibitemShut {NoStop}%
\bibitem [{\citenamefont {Wang}\ \emph {et~al.}(2016)\citenamefont {Wang},
  \citenamefont {Xiong}, \citenamefont {Andrade}, \citenamefont {Venkataraman},
  \citenamefont {Ren}, \citenamefont {Guo},\ and\ \citenamefont
  {Lončar}}]{Wang16}%
  \BibitemOpen
  \bibfield  {author} {\bibinfo {author} {\bibfnamefont {C.}~\bibnamefont
  {Wang}}, \bibinfo {author} {\bibfnamefont {X.}~\bibnamefont {Xiong}},
  \bibinfo {author} {\bibfnamefont {N.}~\bibnamefont {Andrade}}, \bibinfo
  {author} {\bibfnamefont {V.}~\bibnamefont {Venkataraman}}, \bibinfo {author}
  {\bibfnamefont {X.-F.}\ \bibnamefont {Ren}}, \bibinfo {author} {\bibfnamefont
  {G.-C.}\ \bibnamefont {Guo}}, \ and\ \bibinfo {author} {\bibfnamefont
  {M.}~\bibnamefont {Lončar}},\ }\href@noop {} {\bibfield  {journal} {\bibinfo
   {journal} {arXiv preprint arXiv:1610.04197}\ } (\bibinfo {year}
  {2016})}\BibitemShut {NoStop}%
\bibitem [{\citenamefont {Chang}\ \emph {et~al.}(2016)\citenamefont {Chang},
  \citenamefont {Li}, \citenamefont {Volet}, \citenamefont {Wang},
  \citenamefont {Peters},\ and\ \citenamefont {Bowers}}]{Chang16}%
  \BibitemOpen
  \bibfield  {author} {\bibinfo {author} {\bibfnamefont {L.}~\bibnamefont
  {Chang}}, \bibinfo {author} {\bibfnamefont {Y.}~\bibnamefont {Li}}, \bibinfo
  {author} {\bibfnamefont {N.}~\bibnamefont {Volet}}, \bibinfo {author}
  {\bibfnamefont {L.}~\bibnamefont {Wang}}, \bibinfo {author} {\bibfnamefont
  {J.}~\bibnamefont {Peters}}, \ and\ \bibinfo {author} {\bibfnamefont {J.~E.}\
  \bibnamefont {Bowers}},\ }\href {\doibase 10.1364/OPTICA.3.000531} {\bibfield
   {journal} {\bibinfo  {journal} {Optica}\ }\textbf {\bibinfo {volume} {3}},\
  \bibinfo {pages} {531} (\bibinfo {year} {2016})}\BibitemShut {NoStop}%
\bibitem [{\citenamefont {Ilchenko}\ \emph {et~al.}(2004)\citenamefont
  {Ilchenko}, \citenamefont {Savchenkov}, \citenamefont {Matsko},\ and\
  \citenamefont {Maleki}}]{Ilchenko04}%
  \BibitemOpen
  \bibfield  {author} {\bibinfo {author} {\bibfnamefont {V.~S.}\ \bibnamefont
  {Ilchenko}}, \bibinfo {author} {\bibfnamefont {A.~A.}\ \bibnamefont
  {Savchenkov}}, \bibinfo {author} {\bibfnamefont {A.~B.}\ \bibnamefont
  {Matsko}}, \ and\ \bibinfo {author} {\bibfnamefont {L.}~\bibnamefont
  {Maleki}},\ }\href@noop {} {\bibfield  {journal} {\bibinfo  {journal}
  {Physical Review Letters}\ }\textbf {\bibinfo {volume} {92}},\ \bibinfo
  {pages} {043903} (\bibinfo {year} {2004})}\BibitemShut {NoStop}%
\end{thebibliography}%
\end{document}